# Quantum Computing Solution of DC Power Flow

Rozhin Eskandarpour[1], Kumar Ghosh[1], Amin Khodaei[1]*, Liuxi Zhang[2], Aleksi Paaso[2], Shay Bahramirad[2]

[1]Department of Electrical and Computer Engineering, University of Denver, Denver, CO 80210 USA.

[2]Smart Grid and Technology, Commonwealth Edison, Chicago, IL 60618 USA

*Corresponding author: Amin Khodaei, Department of Electrical and Computer Engineering, University of Denver, Denver, CO 80210, USA. Tel: +1-303-871-2481; e-mail: amin.khodaei@du.edu

***Abstract—*** In this paper, we model and solve a fundamental power system problem, i.e., DC power flow, using a practical quantum computer. The Harrow-Hassidim-Lloyd (HHL) quantum algorithm is used to solve the DC power flow problem. The HHL algorithm for the solution of a system of linear equations (SLE) offers an exponential speedup over the classical computers. The speedup advantage is more significant when the size and the frequency of solving the power flow problem become more substantial.

Verified quantum computing applications to power systems are merely nonexistent at this point. Through this paper, we plan to (1) provide a proof-of-concept that practical power system problems can be solved using quantum technology, (2) build a quantum-grid foundation by solving a fundamental power system problem with applications in many other critical grid problems, and (3) apply HHL to solve an SLE which has broad applications in many power system problems. A small 3-bus system is used for testing and demonstration purposes, considering the limitations of the available quantum computing hardware and software. The proposed method's merits and effectiveness are demonstrated using IBM open-source quantum computer and reported through proof-of-concept experimental demonstration using a 4-qubit quantum information processor.

# 1. Introduction

Electric utilities and local governments are dealing with rising expectations of uninterrupted service from electricity customers while experiencing a surge in distribution-connected generation technologies, greater need to rely on renewable energy resources, and growing electrification in various sectors such as transportation. These have made data the epicenter of modern decision making while intensifying data processing challenges, to the point that the incremental improvements in grid analytics and computing cannot handle the changes taking place in the grid today. As a result, substantial progress needs to be made with integrative methods that draw on emerging analytics and computing technologies.

Grid control, operation, and planning modeling involve thousands to millions of variables and equations and require improvements in existing models' efficiency and precision, describing more security and physical impacts while incorporating uncertainty. Therefore, it is imperative to perform parallel information processing and rapid search over unordered sets of data while making predictions and decision-making faster and more accurate. Quantum computing is a promising technology that will help address the current computational challenges and support the journey in building a more resilient, reliable, safe, and secure grid of the future [1].

Quantum computing is developed based on quantum mechanical phenomena that describe the nature and conduct of energy and matter at the level of fundamental subatomic particles. A quantum computer operates by controlling the behavior of these particles to achieve desired computation. Quantum computers mark a step forward in computing capacity, which is potentially superior to a modern supercomputer by offering extensive efficiency. Based on quantum physics laws, a quantum computer can achieve enormous processing power over multi-state capacity and can execute multiple functions simultaneously by using possible permutations [2].

There have been extensive efforts globally to build a quantum computer capable of significantly improving computing power and to support solving problems that classical computers cannot solve [3].

The initial efforts on the quantum technology date back to the 1980s, when it was looked at as a system that can be designed [2]. In 1982, physicist Richard Feynman indicated that quantum mechanical phenomena could be applied to simulate a quantum system more effectively than a naïve simulation on today's classical computers [4][5]. Follow-on notable work included the illustration of violating the extended Church-Turing thesis by quantum computers in 1993 [6], the exponential speedup in quantum polynomial-time algorithms for the discrete logarithm and integer factoring problems in 1994 [7], and exponential acceleration in cryptanalysis through Shor's algorithm also in 1994, thereby potentially threatening several of the commonly-used cryptographic methods [8].

Research and development attempts have produced significant strides in building a working quantum machine in the last two decades. The milestones included the demonstration of basic analog and digital proof-of-principle systems and building the first tiny quantum computer in 1997, and the revealing of a 28-qubit quantum computer by the Canadian startup D-Wave in 2007 [9].

A quantum computer has the theoretical potential to solve some of the problems that no classical computer could currently solve. These are the problems found in many fields, such as finance, cybersecurity and cryptography, forecasting, drug design, molecular modeling, aerospace design, and weather services, to name a few. Quantum computing has swiftly advanced in recent years due to substantial development in algorithms. These advances are carrying quantum computers closer to their impending commercial utility [10][11][12]. Despite these advances, however, the power and energy sector has yet to explore the many advantages of quantum computing.

Building the grid of the future requires applying technologies that can help address unsolved problems and emerging challenges. Many prominent techniques have been created to solve various power system problems, starting from exact enumeration methods [13] to more recent numerical optimization techniques such as stochastic methods. Applications to power systems have been limited to quantum-inspired methods [14][15]. Lacking, however, is the implementation of a practical power system problem in a quantum computer.

This paper contributes to this emerging field of research by solving the DC power flow problem using a quantum computer. DC power flow is an extensively-solved problem in today's power grids due to the vital information it provides. However, the growing size and time-sensitivity of this problem have resulted in extensive efforts to speed up the solution process. This paper focuses on the DC power flow problem and applies the Harrow-Hassidim-Lloyd (HHL) algorithm [16] to find its quantum solution. The rest of the paper is organized as follows. In Section II, an overview of the DC power flow problem along with governing equations is provided. In Section III, the HHL algorithm is reviewed and further applied to the proposed DC power flow problem. Section IV presents the numerical simulation results on a test power system. Section V provides a discussion on the challenges and the next steps of this research, while Section VI concludes the paper.

## 2. DC Power Flow Problem

The reliable operation of a power system depends heavily on several problems that are based on power flow studies. Power flow is a numerical analysis of the power system to determine its current state, under steady-state conditions, based on real and reactive nodal power injections and withdrawals. The solution of the power flow problem includes the flow of electricity on various transmission and distribution branches, line losses, and voltage magnitude and angle on network buses. There is a wide range of problems that are reliant on power flow studies, solved from every few minutes (including control and operation problems) to every year (such as maintenance and planning problems).

The power flow equations follow the system's physics, i.e., how electrons follow the path of least resistance in the network to move from generators to loads. This problem is inherently nonlinear; however, given the need to solve this problem in a timely manner in many of the mentioned problems, there have been extensive efforts to apply reformulation, approximations, or parallel processing to speed up its computation time. DC power flow, which is obtained through a set of approximations, is the most commonly used form of the simplified power flow problem.

Equations (1) and (2) represent real and reactive power flow, respectively.

$$p_m = \sum_n v_m v_n (G_{mn} \cos(\theta_m - \theta_n) + B_{mn} \sin(\theta_m - \theta_n)) \tag{1}$$

$$q_m = \sum_n v_m v_n (G_{mn} \sin(\theta_m - \theta_n) - B_{mn} \cos(\theta_m - \theta_n)) \tag{2}$$

where $m$ is an index for buses, and $mn$ is as an index for lines. $p$ and $q$ are net real and reactive nodal injections, respectively (in which net injection denotes local generation minus local load). $v$ is nodal voltage magnitude, and $\theta$ is nodal voltage angle. $G$ and $B$ represent grid characteristics and are obtained from the network admittance matrix. The DC power flow equations for the transmission network are obtained based on three assumptions:

- In the transmission network, line resistance is much smaller than line reactance, i.e., $r_{mn} \ll x_{mn}$. This is a valid assumption in transmission lines, which will enable discarding conductance from the formulation, where $g_{mn} = r_{mn}/(r_{mn}^2 + x_{mn}^2) \approx 0$.

- The difference in voltage angles of adjacent buses is considered to be small. Thus trigonometric terms for two adjacent buses $m$ and $n$ can be approximated as $\sin(\theta_m - \theta_n) \approx \theta_m - \theta_n$ and $\cos(\theta_m - \theta_n) \approx 0$.

- Voltage magnitudes are assumed to be 1 p.u., i.e., $v_m = v_n = 1$, and as a result, all voltage magnitude and reactive power calculations are ignored.

These three assumptions result in a linear power flow equation, defined as:

$$p_m = \sum_n B_{mn}(\theta_m - \theta_n) \tag{3}$$

$p_m$ are all known in the system based on the power output of generators and forecasted loads. $B_{mn}$ values are also known from the characteristics of the transmission network and calculated from line susceptances ($b_{mn}$) as follows:

$$B_{mn} = -b_{mn} \qquad m \neq n \tag{4}$$

$$B_{mm} = \sum_n b_{mn} \tag{5}$$

The only variables of this problem are $\theta$'s (representing nodal voltage angles) that, once found, can be used to calculate real power flowing on each line. Voltage angles are obtained by solving the system of linear equations (SLE) in (6):

$$B\theta = p \tag{6}$$

DC power flow, as a linear problem that can find an acceptable approximate solution in a fraction of time compared to the full AC power flow, has vast applications in power systems, from real-time security analysis to long-term planning. However, as this problem becomes larger, to address the ever-increasing size of the power networks or integrating a myriad of security scenarios, the classical computers may fail to provide a solution in an acceptable time (as defined by the application), thus justifying the leveraging of quantum computing capabilities.

## 3. Quantum Algorithm for Solving Dc Power Flow Equations

The quantum computer utilizes quantum mechanics to execute certain types of computation more effectively than a classical computer [17][18]. SLE stands up certainly in most real-life applications, such as solving partial differential equations. For the DC power flow, and as discussed in the previous section, the problem can be illustrated as: given a matrix $B \in \mathbb{R}^{N \times N}$ and a vector $p \in \mathbb{R}^N$, find $\theta \in \mathbb{R}^N$ while satisfying $B\theta = p$.

An SLE is called s-sparse if $B$ has at most $s$ nonzero entries per row or column, which is the case for the DC power flow problem. Solving an s-sparse system of size $N$ with a classical computer requires O($N$ $sk$ $log(1/\epsilon)$) running time using the conjugate gradient method [19]. Here $k$ and $\epsilon$ denote the condition number of the system and the approximation accuracy, respectively. The Harrow-Hassidim-Lloyd quantum algorithm, known as the HHL algorithm, is proven to achieve a running time complexity of O($log(N)s^2k^2/\epsilon$) [19], which is a significant speedup compared to its classical counterpart.

## A. Algorithm Description

The problem should be encoded in the quantum language to solve the system of linear DC power flow equations. Initially, $p$ and $\theta$ are normalized and mapped to the respective quantum states $|p\rangle$ and $|\theta\rangle$. The rescaled problem that is to be solved will be as in (7):

$$B|\theta\rangle = |p\rangle \tag{7}$$

Matrix $B$ should be Hermitian so that the HHL algorithm can be applied. This is the case for the DC power flow as this matrix is defined through (4) and (5). It can, therefore, be decomposed as follows:

$$B = \sum_j \lambda_j |u_j\rangle\langle u_j|, \qquad \lambda_j \in \mathbb{R} \tag{8}$$

and accordingly

$$B^{-1} = \sum_j \lambda_j^{-1} |u_j\rangle\langle u_j|, \tag{9}$$

where $|u_j\rangle$ is the $j$th eigenvector of $B$ with eigenvalue $\lambda_j$.

The algorithm exits with the readout register in the following state when the right-hand side is in the eigenbasis of $B$,

$$|p\rangle = \sum_j p_j |u_j\rangle, \tag{10}$$

$$|\theta\rangle = B^{-1}|p\rangle = \sum_j \lambda_j^{-1} p_j |u_j\rangle \tag{11}$$

Note that there is already an implicit normalization constant since a quantum state is discussed in this algorithm.

*B. Algorithm Implementation*

The HHL algorithm is implemented as shown in Fig. 1. The algorithm uses three quantum registers: The first register is used to store a binary representation of the eigenvalues of $B$ and is denoted by $\alpha$. The second register contains the vector solution indicated by $\beta$, where $N = 2^\beta$. The third register is for ancilla qubits. All three registers are set to $|0\rangle$ at the beginning of each computation and restored to $|0\rangle$ at the end.

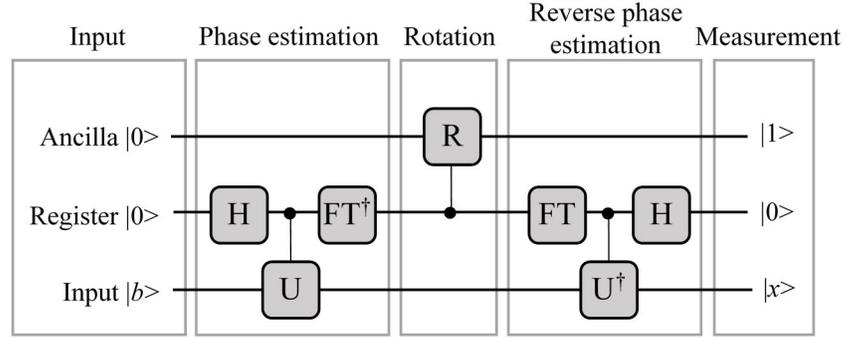

Figure 1. HHL algorithm process

To start the process, initially, the data is loaded into $|p\rangle \in \mathbb{C}^N$, through the following transformation:

$$|0\rangle_\beta \rightarrow |p\rangle_\beta \tag{12}$$

This initial step is followed by HHL's four major steps as described below:

1) Quantum Phase Estimation (QPE). QPE is at the core of the HHL algorithm and finds the eigenvalues (or phases) of an eigenvector of a unitary operator, i.e., a controlled unitary with a change of basis that maps the eigenvalues onto the working memory. Let $U \in \mathbb{C}^{2m \times 2m}$ be unitary and $|\psi\rangle_m \in \mathbb{C}^{2m}$ be one of its eigenvectors with respective eigenvalue $e^{2\pi i \gamma}$. QPE takes as inputs the unitary gate for U and the state $|0\rangle_n |\psi\rangle_m$ and returns the state $|\tilde{\gamma}\rangle_n |\psi\rangle_m$. Here $\tilde{\gamma}$ signifies a binary approximation to $2^n\gamma$, and the $n$ subscript denotes its truncation to $n$ digits.

$$QPE(U, |0\rangle_n|\psi\rangle_m) = |\tilde{\gamma}\rangle_n|\psi\rangle_m \tag{13}$$

For the DC power flow problem, QPE with U=$e^{iBt}$ is used. In this case

$$e^{iBt} = \sum_j e^{i\lambda_j t}|u_j\rangle\langle u_j|, \tag{14}$$

Then, for the eigenvector $|u_j\rangle_\beta$, which has eigenvalue $e^{i\lambda_j t}$, QPE will output $|\tilde{\lambda}_i\rangle_\alpha|u_i\rangle_\beta$, where $\tilde{\lambda}_i$ represents an $\alpha$-bit binary approximation to $2^\alpha\lambda_j t/2\pi$. Therefore, if each $\lambda_j$ can be correctly represented with $\alpha$ bits,

$$QPE(e^{iB2\pi}\sum_j p_j|0\rangle_\alpha|u_j\rangle_\beta) = \sum_j p_j|\lambda_j\rangle\langle u_j| \tag{15}$$

The quantum state of the register stated in the eigenbasis of $B$ is:

$$\sum_j p_j|\lambda_j\rangle_\alpha|u_j\rangle_\alpha \tag{16}$$

where $|\lambda_j\rangle_\alpha$ is the $\alpha$-bit binary representation of $\lambda_j$.

2) Controlled Rotation. An ancilla qubit is added and applies a rotation conditioned on $|\lambda_j\rangle$:

$$\sum_j p_j|\lambda_j\rangle_\alpha|u_j\rangle_\alpha \left(\sqrt{1-\frac{C^2}{\lambda_j^2}}|0\rangle + \frac{C}{\lambda_j}|1\rangle\right) \tag{17}$$

where $C$ is a normalization constant.

3) Uncomputation. This step applied a reverse QPE to uncompute the results of the previous step:

$$\sum_j p_j|0\rangle_\alpha|u_j\rangle_{\alpha,} \left(\sqrt{1-\frac{C^2}{\lambda_j^2}}|0\rangle + \frac{C}{\lambda_j}|1\rangle\right) \tag{18}$$

4) Measurement. This step measures the ancilla qubit. The register is in the post-measurement state when the outcome is 1:

$$\left(\sqrt{\frac{1}{\sum_j |p_j|^2/|\lambda_j|^2}}\right)\sum_j p_j |0\rangle_\alpha |u_j\rangle_\alpha \tag{19}$$

## 4. Numerical Studies

A three-node system, shown in Fig. 2, is used for studying how the HHL algorithm can be employed to solve the DC power flow problem. This system has three nodes (buses) and three lines. The network line characteristics are shown in Table 1. The generation from units at nodes 1 and 2 are at 20 MW and 60 MW, respectively. There is a load of 80 MW at node 3.

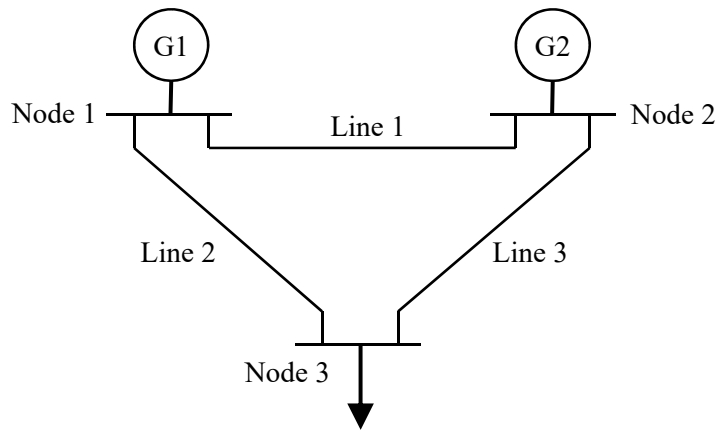

Figure 2. Three-node test system

Table 1. Line characteristics

| Line number | From bus | To bus | Reactance (pu) |
|:---:|:---:|:---:|:---:|
| 1 | 1 | 2 | 0.0125 |
| 2 | 1 | 3 | 0.0125 |
| 3 | 2 | 3 | 0.05 |

*A. HHL Solution Algorithm for DC Power Flow*

The DC power flow equations are represented as in (6). The coefficient matrix is not invertible, so we assume that the voltage angle at node 1 has a value of 0, i.e., $\theta_1 = 0$. This is, in fact, a realistic assumption in power systems as there is always at least one node that acts as the slack bus and provides a reference for nodal voltage angles. By writing the power flow equations and factoring out a constant from both sides, the SLE will be obtained as follows:

$$\begin{bmatrix} 1 & -0.2 \\ -0.2 & 1 \end{bmatrix} \begin{bmatrix} \theta_2 \\ \theta_3 \end{bmatrix} = \begin{bmatrix} 0.6 \\ -0.8 \end{bmatrix}$$

where the matrix of coefficients on the left-hand side represents *B,* and the vector on the right-hand side denotes *p,* as discussed in the previous section.

Consider $\beta$=1 qubit to represent $|p\rangle$ and the solution $|\theta\rangle$, $\alpha$=2 qubits to store the binary representation of the eigenvalues, and one ancilla qubit to save whether the conditioned rotation, hence the algorithm, is successful. For the illustrative purpose of the algorithm, a parameter *t* is selected to rescale the matrix *B*'s eigenvalues during QPE and obtain an accurate binary representation of the rescaled eigenvalues in the $\alpha$-register. In the original article on the HHL algorithm [16], the parameter *t* is represented as the simulation-time for the QPE part.

A calculation of the eigenvalues will give $\lambda_1$= 0.8 and $\lambda_2$=1.2, and the eigenvectors $|u\rangle_1 = \frac{1}{\sqrt{2}}\begin{bmatrix} 1 \\ 1 \end{bmatrix}$ and $|u\rangle_2 = \frac{1}{\sqrt{2}}\begin{bmatrix} 1 \\ -1 \end{bmatrix}$, such that B $= \begin{bmatrix} 1 & -0.2 \\ -0.2 & 1 \end{bmatrix} = \sum_j \lambda_j |u\rangle_j \langle u|_j$

The QPE will output a 2-bit binary approximation to $\frac{\lambda_i t}{2\pi}$. If we set $t= 2\pi.\frac{5}{8}$, then QPE will give a 2-bit binary, where $\frac{\lambda_1 t}{2\pi} = \frac{1}{2}$ and $\frac{\lambda_2 t}{2\pi} = \frac{3}{4}$, which are designated as $|01\rangle_\alpha$ and $|10\rangle_\alpha$. $|p\rangle$ can be accordingly written in the eigenbasis of *B* as $|p\rangle_\beta = 0.6|0\rangle - 0.8|1\rangle = -0.1\sqrt{2}|u_1\rangle + 0.7\sqrt{2}\,|u_2\rangle$. With this initialization, the different steps of the HHL algorithm can be applied. Applying QPE will yield

$-0.1\sqrt{2}|01\rangle|u_1\rangle + 0.7\sqrt{2}|10\rangle|u_2\rangle$

Using (17), the eigenvalues will give conditioned rotation with $C=\frac{5}{8}=0.625$ to compensate from having rescaled:

$$-0.1\sqrt{2}|01\rangle|u_1\rangle\left(\sqrt{1-\frac{(0.625)^2}{(0.8)^2}}|0\rangle+\frac{0.625}{0.8}|1\rangle\right)+0.7\sqrt{2}|10\rangle|u_2\rangle\left(\sqrt{1-\frac{(0.625)^2}{(1.2)^2}}|0\rangle+\frac{0.625}{1.2}|1\rangle\right)$$

$$=-0.1\sqrt{2}|01\rangle|u_1\rangle\left(\sqrt{1-(0.781)^2}|0\rangle+0.781|1\rangle\right)+0.7\sqrt{2}|10\rangle|u_2\rangle\left(\sqrt{1-(0.521)^2}|0\rangle+0.521|1\rangle\right)$$

After applying reverse QPE, from (18), the qubits are in the state, up to a normalization factor $N$:

$$-0.1\sqrt{2}|00\rangle|u_1\rangle\left(\sqrt{1-(0.781)^2}|0\rangle+0.781|1\rangle\right)+0.7\sqrt{2}|00\rangle|u_2\rangle\left(\sqrt{1-(0.521)^2}|0\rangle+0.521|1\rangle\right)$$

On the outcome of 1 when measuring the ancilla qubit, the normalized qubit state in the readout register is:

$$\frac{1}{N}\left(-0.1\sqrt{2}|00\rangle|u_1\rangle(0.781)+0.7\sqrt{2}|00\rangle|u_2\rangle(0.521)\right)$$

where,

$$\frac{1}{N}\left(-0.1\sqrt{2}|u_1\rangle(0.781)+0.7\sqrt{2}|u_2\rangle(0.521)\right)=\frac{|\theta\rangle}{||\theta||},$$

with normalization constant $N=\sqrt{(0.1)^2(\sqrt{2})^2(0.781)^2+(0.7)^2(\sqrt{2})^2(0.521)^2}$, and the state $|\theta\rangle=B^{-1}|p\rangle_\beta$.

Without using extra gates, the norm of the state $|\theta\rangle$ is computed as follows:

$$P[|\text{ancilla}\rangle=|1\rangle]=N^2=||\theta||^2$$

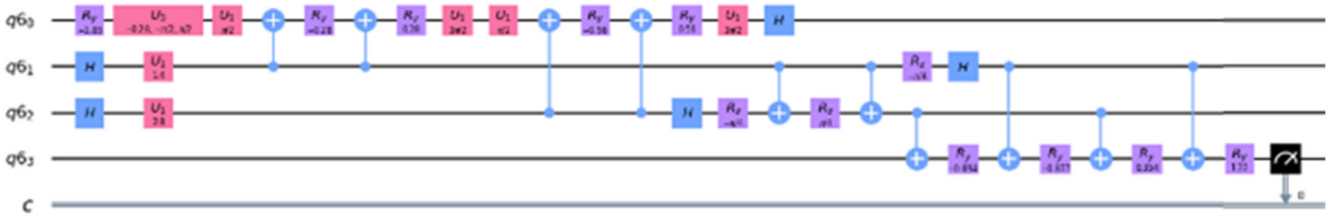

Figure 3. Quantum circuit for HHL algorithm and measurement of ancilla qubit for obtaining $||\theta||$

which is the probability of measuring $|1\rangle$ in the ancilla qubit from the previous step. In the readout register, a normalized quantum state $|\Psi\rangle = \frac{|\theta\rangle}{||\theta||}$ is obtained, which provides an estimate of $||\theta||$ by measuring the ancilla qubit. The estimate of $||\theta||$ is useful because in the end the non-normalized state $|\theta\rangle$ is dealt with for most of the calculations.

## B. Implementation on a Quantum Simulator

The problem is solved analytically in the previous section, and the result of HHL on a quantum simulator is illustrated here. First, a general HHL algorithm provided by Qiskit Aqua is run to obtain an estimate of the fidelity and other parameters. In the DC power flow problem, the time of the Hamiltonian simulation is set to $2\pi.\frac{5}{8}$ . If the matrix has several structures, it might be possible to obtain information about the eigenvalues and use it to choose a suitable $t$ and improve the accuracy of the solution returned by the HHL. The reason for selecting $t$ is to rescale the eigenvalues during QPE, which could be represented accurately with two binary digits. The fidelity of the solution should be near 1 when it is done correctly. Table 2 demonstrates the results of the studied DC power flow problem.

Table 2 DC Power Flow Results

| Fidelity | 0.999945 |
|---|---|
| Probability | 0.020986 |
| Classical Solution | [0.4583  -0.7083] |
| Quantum Solution | [0.4534-0.j  -0.7123+0.j] |

The high fidelity of the results shows the promise in finding accurate solutions. Table II shows the characteristics of the quantum circuit used to solve this problem. The depth is the maximum number of gates applied to a single qubit. The width is the number of qubits required. The number of CNOTs is also provided as this number, and the width, provide a sense of whether running the circuit on current practical hardware is feasible.

Table 3 Characteristics of the Quantum Circuit

| | |
|---|---|
| Circuit Width | 7 |
| Circuit Depth | 102 |
| CNOT Gates | 54 |

*C. Implementation on a Real Quantum Computer*

The above results are obtained from the standard algorithm provided in Qiskit. It can be seen that the above method uses 7 qubits, has a circuit depth 102, and requires 54 CNOT gates. This setup is not feasible for the currently available quantum hardware, so these quantities shall be decreased.

We design a 4-qubit quantum circuit with Qiskit and simulate with Qasm simulator such that after solving the above problem with the HHL method, a normalized vector $|\Psi\rangle = \frac{|\theta\rangle}{||\theta||}$ is obtained in the readout register, such that $|\theta\rangle = ||\theta|| \, |\Psi\rangle$, where the square of the norm $||\theta||^2$ is the probability of measuring $|1\rangle$ in the ancilla qubit.

A parameter $t$=1.4 is selected in the quantum circuit, and the probability of measuring $|1\rangle$ in the ancilla qubit is given by (P(|ancilla⟩=|1⟩) =0.9844)).

Remembering the DC power flow equations, the flow of line $mn$ is calculated as $b_{mn}(\theta_m - \theta_n)$. This is commonly an essential outcome of the power flow problem and one that is used to examine against the line capacity limits. For directly calculating line flows through the quantum algorithm, we estimate $(\theta_m - \theta_n)^2$ and use its square root to calculate line flow.

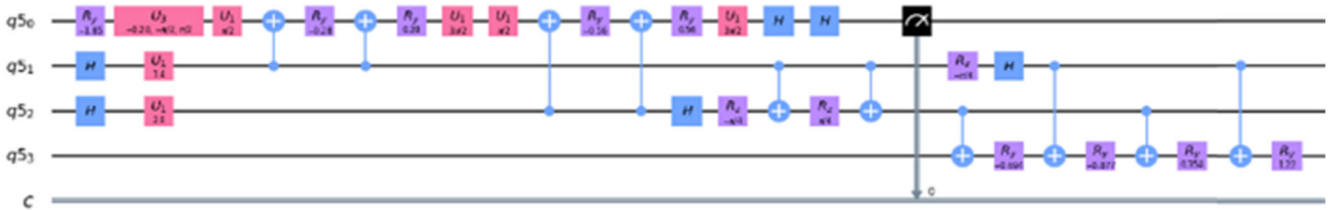

Figure 4. Quantum circuit for calculating the line flow by measuring the readout register.

Assume for the studied test system, flow of line 2-3 is of interest, i.e., $(\theta_2 - \theta_3)^2$ is to be estimated. A Hadamard gate is put at the end of the state qubit such that

$$H|\Psi\rangle = \Psi_1 H|0\rangle + \Psi_2 H|1\rangle = \left(\frac{1}{\sqrt{2}}\right)[(\Psi_1 + \Psi_2)|0\rangle + (\Psi_1 - \Psi_2)|1\rangle].$$

Now,

$$P(H|\Psi\rangle = |1\rangle) = \frac{1}{2}(\Psi_1 - \Psi_2)^2 = \frac{1}{2||\theta||^2}(\theta_2 - \theta_3)^2.$$

Or $(\theta_2 - \theta_3)^2 = 2||\theta||^2 P(H|\Psi\rangle = |1\rangle)$

The corresponding quantum circuit is shown in Fig. 4.

From the exact calculation $\theta_2 = 0.4583$ and $\theta_3 = -0.7083$ such that, $(\theta_2 - \theta_3)_{exact}^2 = 1.3609$. After running the quantum simulation, we get $P(H|\Psi\rangle = |1\rangle) = 0.5999$ and $||\theta||^2 = 0.9844$, such that $(\theta_2 - \theta_3)_{HHL}^2 = 0.5999 \times 2 \times ||\theta||^2 = 1.1810$.

We can see a reasonable agreement for quantum simulation in a noisy environment and statistical errors from the above results. Note that we have used many two-qubit gates (CNOT) in the circuit; in a noisy environment, the two-qubit gates largely affect the simulation's efficiency and accuracy.

## 5. Discussions

This paper focused on solving an extensively-used power system problem with a practical quantum computer to demonstrate the potential benefits of this viable technology in the power and energy sector. There are, however, multiple points to consider:

- The quantum solution was not an exact solution, although it was run on a small test system. The reason is that the existing quantum computers are noisy and not always capable of finding an accurate solution. This challenge is expected to be lessened in the coming years with extensive advances in quantum hardware.

- The quantum speedup was not apparent from the studied test system, given its small size. However, as the problem becomes larger, the quantum advantage becomes more apparent. For example, if the size of this problem is increased by 10,000 times (to represent a more practical network), the computation time in a classical computer will increase by 10,000 times, i.e., $\sim O(N)$ while it would be only 100 times more on a quantum computer, i.e., $\sim O(log(N))$.

- For the practical purpose of designing a quantum circuit, the HHL algorithm has few limitations. For example, in the ideal scenario, to efficiently encode the classical data into a quantum computer, we need a Q-RAM or other efficient quantum algorithm, which is an open problem. Also, for the Hamiltonian simulation (eigenvalue calculation), we need to exponentiate the desired matrix and decompose into a combination of quantum gates, with Trotter-Suzuki decomposition, which itself is a computationally expensive algorithm. This is currently challenging to simulate beyond $8 \times 8$ matrices. Other types of quantum algorithms are popular for eigenvalue calculations and solving systems of linear equations, which are known as Variational Quantum Eigensolver (VQE) [21] and Variational Quantum Linear Solver (VQLS) [22], respectively. However, there is no concrete proof that these algorithms have significant advantages over their classical counterparts.

## 6. Conclusion

The power system is evolving ever more rapidly, not only to embrace new technologies but also to adapt to the changing climate and decrease carbon-based fuels' use to slow down the warming trend. Quantum computers would provide much-needed processing power to the grid of the future. In this paper, the DC power flow problem was solved using a quantum algorithm on a practical quantum computer. A small 3-bus system was used for testing and demonstration purposes, considering the limitations of the available quantum computing hardware and software. The power flow problem for this system was solved analytically, through quantum simulation, and using a practical quantum computer, providing a proof-of-concept that a fundamental power system problem can be solved using quantum technology. As quantum technology progresses, larger test systems can be tested to experimentally illustrate the true quantum speedup.